\documentclass[acmsmall,screen,nonacm,review=false,timestamp=false]{acmart} 

\makeatletter
\def\@ACM@checkaffil{
    \if@ACM@instpresent\else
    \ClassWarningNoLine{\@classname}{No institution present for an affiliation}%
    \fi
    \if@ACM@citypresent\else
    \ClassWarningNoLine{\@classname}{No city present for an affiliation}%
    \fi
    \if@ACM@countrypresent\else
        \ClassWarningNoLine{\@classname}{No country present for an affiliation}%
    \fi
}
\makeatother

\makeatletter
\let\@authorsaddresses\@empty
\makeatother

\AtBeginDocument{%
  }

\setcopyright{acmlicensed}
\copyrightyear{2018}
\acmYear{2018}
\acmDOI{XXXXXXX.XXXXXXX}
\acmConference[Conference acronym 'XX]{Make sure to enter the correct
  conference title from your rights confirmation email}{June 03--05,
  2018}{Woodstock, NY}
\acmISBN{978-1-4503-XXXX-X/2018/06}

\usepackage{csquotes}
\usepackage{graphicx}
\usepackage{multirow}

\begin{document}

\title{In Which Areas of Technical AI Safety Could Geopolitical Rivals Cooperate?}

\author{Ben Bucknall}
\authornote{Equal contribution.}
\email{bucknall@robots.ox.ac.uk}
\affiliation{
    \institution{Department of Engineering Science, University of Oxford \& Oxford Martin AI Governance Initiative, \texttt{bucknall@robots.ox.ac.uk}}
}

\author{Saad Siddiqui}
\authornotemark[1]
\email{saad@saif.org}
\affiliation{%
  \institution{Safe AI Forum \& Oxford Martin AI Governance Initiative, \texttt{saad@saif.org}\vspace{6pt}}
}

\author{Lara Thurnherr}
\affiliation{\institution{King's College London}
}

\author{Conor McGurk}
\affiliation{\institution{Safe AI Forum}
}

\author{Ben Harack}
\affiliation{\institution{Oxford Martin AI Governance Initiative}
}

\author{Anka Reuel}
\affiliation{\institution{Stanford University \& The Belfer Center for Science and International Affairs, Harvard Kennedy School}
}

\author{Patricia Paskov}
\affiliation{\institution{RAND}
}

\author{Casey Mahoney}
\affiliation{\institution{RAND}
}

\author{S\"oren Mindermann}
\affiliation{\institution{Mila - Quebec AI Institute}
}

\author{Scott Singer}
\affiliation{\institution{Carnegie Endowment for International Peace \& Oxford Martin AI Governance Initiative}
}

\author{Vinay Hiremath}
\affiliation{\institution{Centre for the Governance of AI}
}

\author{Charbel-Rapha\"el Segerie}
\affiliation{\institution{Centre pour la S\'ecurit\'e de l'IA (CeSIA)}
}

\author{Oscar Delaney}
\affiliation{\institution{Institute for AI Policy and Strategy}
}

\author{Alessandro Abate}
\affiliation{\institution{Department of Computer Science, University of Oxford}
}

\author{Fazl Barez}
\affiliation{\institution{Department of Engineering Science, University of Oxford}
}

\author{Michael K. Cohen}
\affiliation{\institution{UC Berkeley \& Center for Human Compatible AI}
}

\author{Philip Torr}
\affiliation{\institution{Department of Engineering Science, University of Oxford}
}

\author{Ferenc Husz\'ar}
\affiliation{\institution{University of Cambridge}
}

\author{Anisoara Calinescu}
\affiliation{\institution{Department of Computer Science, University of Oxford}
}

\author{Gabriel Davis Jones}
\affiliation{\institution{Oxford Digital Health Labs, University of Oxford}
}

\author{Yoshua Bengio}
\affiliation{\institution{Mila - Quebec AI Institute}
}

\author{Robert Trager}
\affiliation{\institution{Oxford Martin AI Governance Initiative \& Blavatnik School of Government, University of Oxford}
}

\renewcommand{\shortauthors}{Bucknall, Siddiqui et al.}

\begin{abstract}
  International cooperation is common in AI research, including between geopolitical rivals.
  While many experts advocate for greater international cooperation on AI safety to address shared global risks, some view cooperation on AI with suspicion, arguing that it can pose unacceptable risks to national security.
  However, the extent to which cooperation on AI safety poses such risks, as well as provides benefits, depends on the specific area of cooperation.
  In this paper, we consider technical factors that impact the risks of international cooperation on AI safety research, focusing on the degree to which such cooperation can advance dangerous capabilities, result in the sharing of sensitive information, or provide opportunities for harm.
  We begin by why nations historically cooperate on strategic technologies and analyse current US-China cooperation in AI as a case study.
  We further argue that existing frameworks for managing associated risks can be supplemented  with consideration of key risks specific to cooperation on technical AI safety research. 
  Through our analysis, we 
  find that research into AI verification mechanisms and shared protocols may be suitable areas for such cooperation.
  Through this analysis we aim to help researchers and governments identify and mitigate the risks of international cooperation on AI safety research, so that the benefits of cooperation can be fully realised.
\end{abstract}

\maketitle

\section{Introduction}
\begin{displayquote}
\textit{``Many risks arising from AI are inherently international in nature, and so are best addressed through international cooperation.''} -- Bletchley Declaration, 2023 \cite{department_for_science_innovation_and_technology_bletchley_2023}
\end{displayquote}

International cooperation has long been a part of managing risks from advanced technologies. During the Cold War, despite intense rivalry, the United States and Soviet Union collaborated on nuclear verification methods through initiatives like the Joint Verification Experiment, which facilitated progress on arms control agreements \cite{center_for_arms_control_and_non-proliferation_fact_2017}. Today we observe ongoing international cooperation in artificial intelligence (AI) development, including between geopolitical rivals such as the US and China. Additionally, recent years have seen numerous high-level calls for international cooperation specifically on AI safety and governance, from the landmark agreement at the Bletchley AI Safety Summit in 2023 \cite{department_for_science_innovation_and_technology_bletchley_2023} to the consensus statements issued by the International Dialogues on AI Safety\footnote{\url{https://idais.ai/}} \cite[see also,][]{singer_how_2024}.

As in historical analogues, cooperating on the safety of advanced geopolitically sensitive technologies such as AI could play an important role in managing emerging risks. However, cooperation between geopolitical rivals carries its own risks that can and should be carefully weighed in order to ensure that the benefits of cooperation can be fully realised by all parties. This is no less true for safety research on AI (`technical AI safety'). Some AI safety techniques have `capability externalities', as improvements in safety concurrently provide gains in model performance -- for example, reinforcement learning from human feedback (RLHF), as discussed in \cite{dai_safe_2023}. Furthermore, some areas of AI safety such as model evaluations for chemical, biological, radiological and nuclear (CBRN) capabilities involve sensitive national security-related information which could be leaked to cooperators. The process of cooperation may also provide avenues for motivated actors to cause harm, such as by placing backdoors in jointly developed infrastructure. Many frameworks and risk management processes from governments exist to guide international technology cooperation at a general level, such as suggesting researchers conduct due diligence on the identity of their counterparties. However, as leading governments' and researchers' continued focus on deepening the science of AI safety demonstrates \cite{bengio_international_2025, us_department_of_commerce_us_2024}, additional analysis is needed to fully address the specific technical characteristics and geopolitical risks associated with cooperation on AI safety research.

This paper addresses this gap by identifying technical risks specific to cooperation on AI safety research, focusing on the impact of cooperation on advancing capabilities, sharing sensitive information and providing opportunities for harm.
We analyse the relative risk of cooperation for four different areas of technical AI safety 
and assess the feasibility of proposed cooperation in each area.
We do not aim to definitively identify the most suitable areas for cooperation, nor to investigate specific benefits of cooperation in different AI safety research areas, though believe these would be valuable directions for future work.

This paper is roughly broken into three parts. The first provides an overview of cooperation on strategic technologies, through i) outlining why geopolitical rivals typically cooperate on strategic technologies; ii) surveying the existing state of cooperation between rivals on AI research; and iii) looking at how risks from cooperation are currently managed.
The second part highlights four potential risks to which actors cooperating with rivals on topics in technical AI safety may be exposed. 
The final part builds on this by assessing the extent to which proposed areas of cooperation on technical AI safety may succumb to these identified risks.
We give a brief overview of each area, as well as a discussion regarding its suitability for cooperation in light of the potential risks previously introduced, finding that research into i) AI verification mechanisms and  ii) protocols may be areas that are particularly well-suited for international cooperation.

\subsection{Key concepts and definitions}
\subsubsection{Coordination, collaboration, and cooperation}
For the purpose of this paper, we adopt the following definitions.\footnote{There is limited academic consensus on what the terms coordination, cooperation, and collaboration refer to. Work in international relations and meta-analyses of usage in management science arrive at different conclusions \cite{castaner_collaboration_2020,stein_coordination_1982}.}
\begin{itemize}
    \item We use \textbf{coordination} to refer to multiple parties acknowledging shared viewpoints and/or agreeing to pursue common goals independently and in parallel.
    \item We use \textbf{collaboration} to refer to multiple parties jointly pursuing shared goals.
    \item We use \textbf{cooperation} to refer to the overarching category of both coordination and collaboration.
\end{itemize}
As demonstrative examples, we regard the Seoul Ministerial Statement, signed by ministers from 27 nations and the EU, as a case of \textbf{coordination}. It, among other things, \textit{`reaffirm[ed] [the signatories'] shared intent to guide the design, development, deployment, and use of AI in a manner which harnesses its benefits for good'} though does not represent a joint effort to act on this intent \cite{department_for_science_innovation_and_technology_seoul_2024}. In contrast, the US and UK AI Safety Institutes conducting a \textit{`joint safety research and testing exercise'} \cite[as in][]{noauthor_us_2024} is a case of \textbf{collaboration}, as it represents a joint activity undertaken to achieve a shared goal. Both are examples of \textbf{cooperation}.

\subsubsection{Technical cooperation}
In this paper, we limit our scope to considering cooperation on areas of technical AI safety and governance -- that is, research and development within computer science, engineering, mathematics (or similar) aimed at advancing the safety of AI systems, or methods for their effective governance, respectively.\footnote{For overviews of AI safety and technical AI governance, see \cite{hendrycks_introduction_2024} and \cite{reuel_open_2024}, respectively.} Taken together, activities included in the present scope include, but are not limited to: academic scientific research; industry research and development on AI systems or related technologies, such as semiconductors; and state-state collaboration on technical topics, for example through their respective AI Safety Institutes. Prior work has explored potential areas for international collaboration, including on non-technical topics such as the sharing of \textit{`best practices and lessons learned from AI regulation efforts in specific countries'} \cite{guest_topics_2024}.

\section{Background and motivation}
In this section, we provide a historical overview of cooperation on strategic technologies. We outline why geopolitical rivals typically cooperate on strategic technologies, before surveying the existing state of cooperation between rivals on AI research, focusing on the US and China as a case study. Finally, we discuss how risks from cooperation are currently managed,     
serving as background for the section that follows, in which we propose a framework to consider the specific risks from cooperation on AI safety research.

\subsection{Why do geopolitical rivals cooperate on strategic technology?}
Here, we draw on basic concepts from game theoretic accounts of international cooperation and competition \cite[e.g.][]{coe_why_2020,jervis_cooperation_1978} to describe a subset of the plausible reasons geopolitical rivals might consider it to be in their interest to cooperate on strategic technologies such as AI. We illustrate these concepts with empirical examples cited in the literature about AI safety cooperation that we consulted over the course of our research.

First, rivals may cooperate to manage risks from technology that cannot be effectively managed by any single actor. This includes cooperation on risks that cross borders, such as illicit use of technology by international criminal groups. Recent examples of cooperation in this vein include Sino-American agreements to jointly combat money laundering through cryptocurrency \cite{sun_us_2024}. Rivals may also cooperate where collective action is necessary to reduce risk, such as in the November 2024 agreement between the US and China to maintain human control over the decision to use nuclear weapons and avoid integrating AI into nuclear command and control systems \cite{renshaw_biden_2024}.

An actor leading on a given technology may also cooperate 
by unilaterally sharing technology, if doing so is also in the leader's own interests and sharing is  technologically feasible. In the early 1960s, the US shared basic designs for Permissive Action Links (PALs) -- devices that prevent unauthorised nuclear detonation -- with the Soviet Union. This cooperation was made possible by the mutual recognition of clear benefits in preventing accidental escalation, and that early PALs were simple enough to explain without compromising sensitive weapons information \cite{ding_keep_2024}.

Rivals may cooperate to improve geopolitical stability by creating mechanisms that reduce uncertainty and the risk of unintended escalation. Examples here include the Open Skies Treaty, which allows participating nations to conduct unarmed aerial surveillance flights over each other's territories using standardised sensor technologies, creating predictable patterns of interaction between rival militaries, and establishing technical protocols for verification \cite{arms_control_association_open_2021}.\footnote{As part of measures to ensure the security of the observed country, the treaty gives the observed country the right to conduct a pre-flight inspection of the observation aircraft to ensure it is unarmed and equipped only with the agreed-upon sensors with pre-determined resolutions. The observed country can also supply its own observation aircraft for the other country to use \cite{arms_control_association_open_2021}.}

Finally, rivals may also cooperate to pool expertise and resources when technological development costs exceed the resources or capacity of any single actor. The International Space Station partnership between space agencies, including those of the U.S. and Russia, leveraged each state's space capabilities \cite{mauduit_collaboration_2017}, while the ITER fusion project brings together rival powers to share the massive costs and technical challenges of developing fusion power.\footnote{See \url{https://www.iter.org/few-lines}. Despite pooling of resources, ITER is 9 years behind schedule, with private fusion startups on track to outpace it \cite{clery_giant_2024}.}

The reasons specified above hold true for states, but at the academic and corporate level, cooperation can take place for reasons completely unrelated to broader geopolitical tension. For example, academics may collaborate with colleagues based in a rival nation who have valuable expertise in a particular area. Companies may want to cooperate with other companies based in rival jurisdictions because they fulfill complementary roles in a complex globalised supply chain, or to ensure interoperability of products to retain market access in key markets.

Many of these reasons for cooperation likely also apply to AI and AI safety. For example, increasingly capable AI systems developed in one jurisdiction may have negative cross-border impacts in a rival jurisdiction, requiring cooperation between rivals to manage these risks effectively. AI safety techniques developed by one state that reliably reduce the risk of an AI system self-replicating and self-improving without human approval could be shared if they reduce overall global risk, and do not reveal sensitive commercial or national security-relevant information.

\subsection{Cooperation on AI case study: The US and China}
In this section we outline the current state of international cooperation within the realms of academia, industry, and intergovernmental relations, focused on the two leading global AI powers, the US and China.\footnote{While we focus on cooperation within each level of academia, industry, or state, it is worth noting that much cooperation can also take place between these communities, in the form of academic-industry partnerships, or academic participation in policy processes, as demonstrated, for example, in the members of the UN's High-Level Advisory Body on AI (see \url{https://www.un.org/en/ai-advisory-body/members}).} The picture that emerges is one where significant academic and some industry cooperation continue to take place, even given the limited cooperation between the respective governments.

\subsubsection{Academia}
US and Chinese researchers collaborate more than researchers from any other pair of countries \cite[][\S1.1]{zhang_ai_2022}, including on topics in AI safety -- as shown in Figure \ref{fig:UScoauthor} below. China surpassed the UK as the largest collaborator with American researchers in 2017, and has retained this position since, although a significant majority of AI safety research is conducted by American researchers without international co-authors.

\begin{figure}
    \centering
    \includegraphics[width=0.75\linewidth]{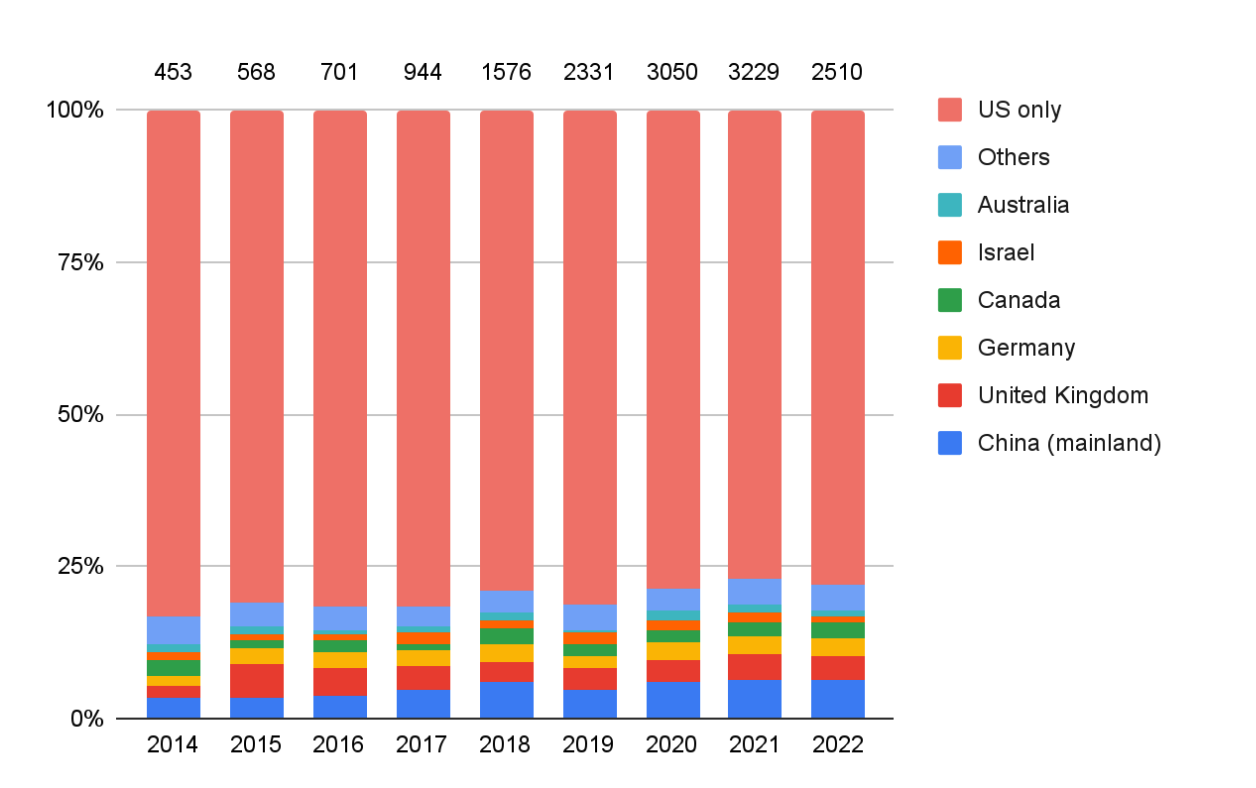}
    \caption{AI safety co-authorship instances with American researchers (\%). Incomplete data from 2023 and 2024 excluded. Data labels at the top of bars show the total number of AI safety papers by US researchers in that year.  Data source: \cite{emerging_technology_observatory_country_2024}. (Note that if a publication lists authors from organisations in more than one country (excluding the US), the publication will ``count towards'' multiple countries, and thus be represented multiple times in the figure.)}
    \label{fig:UScoauthor}
\end{figure}

\subsubsection{Industry}
Historically, some American firms have set up localised joint ventures in China to establish a foothold in Chinese markets and draw on talent pools. At the same time, these ventures also provided a significant boost to the Chinese technology industry.\footnote{In certain industries (e.g., insurance), foreign companies are required to establish a joint venture with local Chinese partners. In a more restricted set of industries such as aviation, foreign ownership is also limited to below 50\% \cite{oconnor_how_2019}.} Recent authoritative histories reveals how the growth of the Chinese semiconductor industry since the late twentieth century serves as a prime example of this dynamic \cite{miller_chip_2023}. The most prominent example in the AI software sector is Microsoft Research Asia (MSRA), a research institution set up by Bill Gates in 1998 and first led by Kaifu Lee, founder of 01.AI, a leading Chinese open-source AI company \cite{jingjing_whampoa_2016}. These investments in Chinese AI ventures have had an impact on the global AI industry. For example, in 2015 a team from MSRA led by Kaiming He introduced `deep residual nets' \cite{he_deep_2015} and greatly advanced the frontier of deep learning methods.

\subsubsection{Intergovernmental}
Cooperative efforts on AI between the American and Chinese governments have been much more limited than in the case of industry or academia.\footnote{Cooperation on AI has also been affected by American export controls on semiconductors, explicitly designed to safeguard an American advantage in AI by restricting Chinese access to cutting-edge semiconductors required for advanced AI \cite{benson_updated_2023}.} AI has only recently come under the spotlight as an important geopolitical issue, requiring the attention of senior government officials. For example, only in 2023 was AI included as a summit-level topic for a meeting between US President Joe Biden and Chinese President Xi Jinping \cite{huang_three_2023}. In 2024, a dedicated intergovernmental dialogue on AI took place in Geneva in May, with further dialogues planned for the future \cite{martina_us_2024}. Notably, in November 2024, the two heads of state reached an agreement to maintain human control over the decision to use nuclear weapons and avoid integrating AI into nuclear command and control systems \cite{renshaw_biden_2024}.

\subsection{How are risks from cooperation managed?}
The above case study illustrates that there is significant cooperation ongoing between rivals. This section draws on authors' engagement with AI expert communities to provide an illustrative overview of the commonly cited measures used to manage the risks arising from such cooperation generally.

States have been cognizant of the risks of cooperation on strategic technology, and have put in place measures to address such risks.\footnote{It is worth noting that measures that may appear to be sufficient at a time when a technology is not deemed sensitive (e.g., AI in the early 2000s, nuclear research prior to its securitisation), may no longer be sufficient when the national security risks related to the technology become more evident \cite[see e.g.,][]{center_tech-politik_2019}. As such, in this section, we do not take a position on whether existing measures are sufficient.} However, the risk management processes related to cooperating with a rival at the inter-state level are opaque.

More detailed public guidelines exist for academics and companies who, as noted above, may want to cooperate for reasons largely unrelated to geopolitical tensions and may be less aware of relevant geopolitical considerations.

Academics who cooperate with other academics in rival jurisdictions typically must engage with national guidance systems. These tend to require actions such as assessing the risks related to the subject or domain of research (e.g., whether knowledge produced could be misused) as well as the research conditions (e.g., potential military links of collaborators or the political environment of where a collaborator is based) \cite{noauthor_dealing_2023,ukri_managing_2022}. They may also be asked to check the identity of collaborators and their institutions against sanctions lists, and to consult documents such as the Bureau of Industry and Security's (BIS) Commerce Control List, which lists a series of items (e.g., encryption software) that are subject to export control regulations \cite{hirschhorn_us_2022}.
 
For companies engaging in activities such as joint ventures, additional rules on outbound and inbound investments typically apply. In the US, for example, companies must report relevant inbound transactions to investment screening entities, such as the Committee on Foreign Investment in the United States (CFIUS), which reviews foreign investments that may impact US national security \cite{cimino-isaacs_committee_2024}.

One gap in the risk management process outlined above is a conceptual one -- few of the tools above focus on technology-specific nuances. Therefore, actors considering cooperation lack a clear framework to assess how cooperation on their specific technology of interest could pose geopolitically-relevant risks.

\section{Risks of cooperation on AI safety}
In this section we outline risks associated with international cooperation that are specific or particularly pertinent to AI safety. Namely, we outline risks associated with the advancement of (potentially harmful) AI capabilities, exposure of sensitive information regarding strategic technology, and opportunities for motivated actors to take harmful actions that may be afforded by cooperation. As above, we draw on basic concepts from game theoretic accounts of international cooperation and competition to typologise the risks geopolitical rivals might articulate as barriers to cooperation.\footnote{We make no claim that these categories of risks are comprehensive, rather, that they are particularly relevant in the context of cooperating on AI safety.}

\textbf{Developments in AI safety could advance the global capabilities frontier.}
Geopolitical rivals may be hesitant to cooperate on AI safety due to the risk that doing so may, as a side effect, advance the global frontier of (potentially harmful) AI capabilities \cite{guest_safeguarding_2023}. This could as a result of correlations between the safety and capability of AI systems, or if advances made in AI safety can be reapplied or repurposed to increase a system's suitability for deployment \citep[see e.g.,][pg. 29]{hendrycks_overview_2023}. Inadvertently increasing the capabilities of systems through cooperative safety research would be undesirable to an actor that would have a preference for being the sole beneficiary of such capabilities gains. It is worth also noting that the potential of advancing overall dual-use capabilities could be an argument for not pursuing the research, even unilaterally, depending on the specific research project in question, and the actor's risk appetite.


\textbf{Cooperation could differentially advance a rival's strategic AI capabilities.}
Alternatively, a state that is (or is perceived to be) `leading' in terms of strategic AI capabilities, may be unwilling to cooperate with a rival due to a risk that doing so will allow the rival to improve their capabilities relative to the leader's. Such a differential benefit to a rival could be gained through application of the results of cooperative research, or as a side effect of increased access to knowledge or resources provided for the purpose of the collaboration.

\textbf{Cooperation could expose sensitive information regarding nationally strategic technology.}
Cooperation on AI safety could also be risky if the specific focus of the cooperation intersects with other (non-AI) nationally strategic technology, in ways that may raise national security concerns. For example, research that aims to conduct risk modeling to estimate the risk posed to domestic digital infrastructure by AI's potential cyber-offensive capabilities may require in-depth knowledge of this infrastructure. Thus, a meaningful cooperation on such a project may require disclosing such information to a rival -- something that may not be within the risk appetite of many geopolitical actors.

\textbf{Cooperation on AI safety may present opportunities for motivated actors to cause harm.}
Finally, cooperation on AI safety research could allow rivals opportunities to take harmful or malicious action, in cases in which they may be inclined to do so. This could be, for example, through inserting backdoors into systems to which collaborations have access for the purpose of the collaboration, or through misusing resources shared for the purpose of cooperation.

\section{Candidate areas of cooperation}
\label{sec:candidate_areas}
In this section we present a non-comprehensive overview of four areas of AI technical safety -- verification, protocols, infrastructure, and evaluation -- on which international cooperation is emerging, or otherwise has been prominently advocated for. For each area we aim to assess the extent to which cooperation in that area would carry the above risks. A summary of these assessments is given in Table \ref{tab:Candidate_areas}, and we provide examples of ongoing or proposed cooperation in each area in the list below. 

\begin{itemize}
    \item \textbf{Verification mechanisms.} Research into potential methods for verifying the veracity of claims made about systems was highlighted as a potential area of cooperation at a recent track-II dialogue between Western and Chinese academics.\footnote{See \url{https://idais.ai/dialogue/idais-venice/}.} See also, e.g., \citep[][pg. 207]{bengio_international_2025}.
    \item \textbf{Protocols.} Development of codified protocols and best practices, for example `AI safety frameworks' or standardisation efforts, have been the focus of ongoing international cooperation, for example, in the Frontier AI Safety Commitments \cite{department_for_science_innovation_and_technology_frontier_2024}, and the subsequent `Conference on Frontier AI Safety Frameworks' hosted by the UK AI Safety Institute \cite{uk_ai_safety_institute_conference_2024}.
    \item \textbf{Infrastructure.} Developing shared AI infrastructure, or otherwise methods for distributing access to existing infrastructure is called for in the UN High-Level Advisory Body's final report \cite{un_hlab_govering_2024}. Furthermore, the International Network of AI Safety Institutes used evaluation infrastructure developed by two constituent members (\textit{Inspect} and \textit{Moonshot}) when conducting a pilot testing exercise \cite{international_network_of_ai_safety_institutes_improving_2024}, though these platforms are not themselves the result of international cooperation.
    \item \textbf{Evaluations.} AI evaluation and testing exercises have been the subject of ongoing international collaboration between the UK and US AI safety institutes \citep[see e.g.,][]{noauthor_us_2024, international_network_of_ai_safety_institutes_improving_2024}.
\end{itemize}

Based on our assessment, we judge research on verification mechanisms and protocols to be less-challenging areas for international cooperation than infrastructure and evaluations. However, this does not imply that efforts to cooperate on the latter two areas should be avoided.



\subsection{Research on verification mechanisms}
Verification mechanisms are technical procedures that could allow for the certification of claims about an AI system or related resources (for example, the locations and specifications of data centers) \cite{brundage_toward_2020,dalrymple_towards_2024}.\footnote{For related discussion regarding the geopolitics of verification in the case of arms control, see \cite{gallagher_politics_1997} and \cite{coe_why_2020}.} Note that we are here using the term `verification' in a somewhat different sense than the established technical meaning of ensuring that a software or hardware system possesses certain properties through, for example, formal methods such as model checking \cite{baier_principles_2008}. While this is included in our use of the term (see `formal verification' below), we do not restrict ourselves to this definition, instead, including activities such as verifying compliance to international agreements between state actors, be it regarding models, or another part of the AI value chain.

While related, this is distinct from designing a new system, or from uncovering information about a system (for example through conducting evaluations) as it predominantly only strengthens existing knowledge claims made about a system or activity.\footnote{An example from \cite{reuel_open_2024}: \textit{`an assessment task could be to uncover details about the data that a given model was trained on. In contrast, a verification problem could be, given a dataset and a model, to confirm or refute the claim that the model was trained on the dataset.'}} The process of developing verification mechanisms, however, could allow rivals to gather sensitive information (e.g., inspectors in data centers could gather detailed information about installed hardware and its vulnerabilities).

Cooperation between multiple actors on the development of verification mechanisms could be beneficial for advancing mutual trust in any resulting solutions, and therefore increasing the chances that they are applied in practice. Trusted international agreements will likely depend on establishing verifiability throughout the entire technical and procedural stack of dependencies -- cooperation from as far down this stack as possible could help establish mutual verifiability. Furthermore, cooperation could aid in ensuring interoperability or compatibility of verification mechanisms, enabling their use alongside each actor's existing technology and infrastructure with minimal modifications required.

\subsubsection{Risks of cooperation}

\textbf{Advances global frontier capabilities.}
To the extent that some applications of verification concern the attestation of properties of a system, rather than demonstrating the existence of such properties, research in such areas is unlikely to advance the capabilities of AI systems. For example, developing methods for attesting to the amount of compute used in a given training run likely does not bear on frontier AI capabilities. However, this may not hold for all areas of verification. For example, the use of formal verification methods such as model checking could uncover new system properties that were not previously known to developers \cite{baier_principles_2008}.

\textbf{Differentially advances a rival's capabilities.}
As noted above, the development and use of some verification mechanisms could uncover previously-unknown model properties. This could result in a differential advancement of a rival's strategic capabilities. However, there are also many areas that likely avoid this risk, as the claims and information being verified can be restricted to those already known to all parties. As an analogue, in the Open Skies Treaty, countries agree to allow other countries to carry out observation flights over their territory, but only using certified sensors with a specific predetermined resolution \cite{arms_control_association_open_2021}.

\textbf{Exposes other sensitive information.}
A concern associated with cooperating on the development of verification technologies for AI could be if doing so requires disclosing sensitive information about each actor's existing technologies. For example, developing hardware mechanisms for verification \cite{kulp_hardware-enabled_2024} could require deep knowledge of existing hardware technology, which may be viewed as sensitive information that should not be shared with rivals. Some verification procedures, for example allowing inspectors in AI data centers to certify particular claims about the amount of compute present, could result in sensitive information related to or adjacent to the object of verification, such as the types of security systems installed, to be inadvertently disclosed. Some proposals for hardware-enabled verification mechanisms could potentially bypass some of such concerns through being independent of the specification of the AI accelerator and other related hardware being used for training or inference \cite{petrie_interim_2024}.

\textbf{Provides opportunity for motivated actors to take harmful action.}
Jointly developing mechanisms for verification could allow motivated rivals to covertly insert `verification backdoors', allowing them to feign compliance in cases in which this mechanism is applied. The level of difficulty of doing so without the backdoor being discovered would likely vary greatly depending on the specific verification mechanism. This risk could be managed by developing the verification stack in the open -- for example, open-source with significant bug bounties -- as a way of increasing the chances that such subversive attempts would be discovered.

\subsubsection{Subareas}

\textbf{Formal verification.}
Formal verification refers to methods for proving that an AI system meets a specification related to the context and environment it is deployed in \cite{dalrymple_safeguarded_2024}. While formal verification for software has a decades-long history \cite{hoare_axiomatic_1969,naur_proof_1966}, most existing methods struggle with the complexity of massive software systems deployed in the real world \cite{leveson_certification_2023}. More recently, the accelerating inclusion of machine learning models in safety-critical systems has spurred new research extending formal methods to ML \cite[see e.g.,][]{urban_review_2021,boudardara_review_2024}. However, these methods are severely underdeveloped and inadequate for addressing open-ended safety-critical scenarios, language models, or complex autonomous agents interacting with humans.

\textbf{Verifiable audits.}
Cooperation could take place on developing and implementing methods for conducting verifiable audits -- that is, audits for which reported results can be verified as being accurate \cite[][\S5.4.1]{garg_experimenting_2023,south_verifiable_2024,reuel_open_2024}. This could extend to verifying whether a given AI model was trained on a given dataset \cite{choi_tools_2023}. In addition to being an area whose development could be the subject of international cooperation, such methods could be broadly beneficial for building trust in the reported capabilities of a state's own AI systems.

\textbf{Compute usage.}
Verifying the use of specialised computational hardware could be instrumental for checking compliance with international agreements relating to AI \cite{shavit_what_2023,sastry_computing_2024}. While some proposals for secure hardware verification have been proposed \cite{petrie_interim_2024, kulp_hardware-enabled_2024} the development of such methods is nascent. Furthermore, compute verification mechanisms would likely need to be robust to physical tampering so as to disincentivise subversion attempts \cite{aarne_secure_2024} or incorporate other verification mechanisms (e.g., spot checks of specific chip shipments) as fail-safes.

\textbf{Generated content.}
Methods for reliably identifying and verifying AI-generated media as such, for example, through embedding machine-readable watermarks in system outputs, is an ongoing area of research \cite{witt_perfectly_2023, dathathri_scalable_2024, ghosal_survey_2023}, and has gained traction as a potential regulatory requirement.\footnote{See for example, Recital 133 of the EU AI Act \cite{noauthor_regulation_nodate}.} Verification of AI-generated media may also be able to be achieved through cryptographically bound provenance metadata, for example as detailed by the C2PA specification.\footnote{See \url{https://c2pa.org/specifications/specifications/2.1/index.html}.}

\subsection{Codification of protocols and best practices}
Protocols and best practices refer to codified statements of procedures for attaining positive outcomes from AI research and development. Protocols could specify procedures to be followed, or outcomes to be achieved, either by industry or state actors. Developing such shared protocols could be a suitable area for intergovernmental coordination due to its less technical nature, especially if phrased in terms of outcomes rather than actions \cite{schuett_principles_2024}. While some states are already coordinating on the development protocols (for example, through the G7 Hiroshima Process\footnote{\url{https://www.soumu.go.jp/hiroshimaaiprocess/en/index.html}}) there is scope for broadening the extent of cooperation, as well increasing the level of detail at which protocols are specified.

It should be noted that protocols are voluntary or binding codifications of existing knowledge, not purely academic research. As such, they could be more politicised than the other categories discussed here -- especially in international settings. This could bring with it both positives, for example, leveraging technical research to build real world political agreement, as well as negatives, such as if the process becomes co-opted and overly politicised, leading to a degradation in the scientific rigour of the process and outputs.\footnote{See \cite{krasner_global_1991} for a canonical treatment of this dynamic, or \cite{yates_engineering_2019} for a more recent international history.}

\subsubsection{Risks of cooperation}

\textbf{Advances global frontier capabilities.}
Developing protocols may be particularly suited for cooperation between state actors, especially on more established topics for which the protocols aim to codify existing techniques and knowledge. In such areas, protocols would be more a process of standardisation rather than advancing state of the art research, and would thus not be at risk of advancing global AI capabilities.

\textbf{Differentially advances a rival's capabilities.}
To the extent that cooperating on the codification of protocols focusing on areas where all parties have shared knowledge and understanding, the risk of differentially advancing a rival's capabilities through such cooperation is minimal.

\textbf{Exposes other sensitive information.}
As for the previous risk, due to developing protocols being aimed at codifying mutual knowledge in a structured framework that can be agreed upon by multiple actors, no sensitive or private information would necessarily need to be shared with a rival. While this may depend on both the focus area, as well as the level of specificity of the resulting protocols (for example, detailed best practices regarding national security), such areas could be avoided while continuing to cooperate on less sensitive topics.

\textbf{Provides opportunity for motivated actors to take harmful action.}
Given that the codification of protocols does not involve direct involvement with AI systems, cooperating on such codification would not allow rivals to take directly harmful actions. However, prior examples in standardisation suggest that both states and industry actors tend to use the international standardisation process to advance their own interests, potentially to the detriment of other actors \cite{krasner_global_1991, ruhlig_geopolitics_2023, zuniga_geopolitics_2024}. For example, the World Trade Organisation's \emph{`Technical Barriers to Trade'} (TBT)\footnote{\url{https://www.wto.org/english/tratop_e/tbt_e/tbt_e.htm}} agreement requires that countries use international standards instead of domestic ones to prevent favoritism. However, in practice this provides cover for Chinese actors to enforce ITU standards, largely pushed through by and advantageous to Chinese actors, on international companies in China. Nonetheless, misuse of standards-setting processes from Chinese actors is more the exception than the rule and the most influential standards-setting bodies continue to function effectively \cite{sheehan_what_2023}.

\subsubsection{Subareas}

\textbf{Safety frameworks.}
The past year has seen a number of frontier AI developers write and publish safety frameworks \cite{google_deepmind_frontier_2024, openai_preparedness_2023, anthropic_responsible_2024}, with 13 further developers agreeing to publish their own documents by the French AI Action Summit, as part of the Frontier AI Safety Commitments \cite{department_for_science_innovation_and_technology_frontier_2024}. The Chinese Academy of Communication and Information, a think-tank housed under the Ministry of Industry and Information in China, has also released a set of safety-focused voluntary commitments signed by several leading Chinese AI developers \cite{caict_protecting_2024}. While such frameworks to-date have been developed within industry, with the Seoul commitments also being directed towards and signed solely by industry actors, there is a role to play in the public sector in setting the criteria by which these frameworks are assessed, and ensuring that industry actors fulfil their commitments \cite{fli_fli_2024, alaga_grading_2024}. Such cooperation has precedent in the form of many prior national and international frameworks regarding the responsible development and use of AI \cite[see e.g.,][]{national_institute_of_standards_and_technology_us_artificial_2024, the_white_house_blueprint_2022, world_health_organization_ethics_2021}.

In particular, multilateral cooperation could aim to develop and define best practices regarding any of the three outcomes specified in the Seoul commitments (or otherwise) -- for example, advancing methods for identifying, assessing, and managing risks associated with AI development and deployment, or determining thresholds for when models might pose unacceptable levels of risk \cite{koessler_risk_2024}, and what should be done in case such predefined thresholds are breached \cite{karnofsky_if-then_2024}.

\textbf{AI incident standards.}
Actively monitoring for incidents, and having standardised methods for incident reporting, can inform policymakers about harms caused by AI systems \cite{stein_safe_2024, stein_role_2024}. While existing AI incident databases\footnote{See, for example, the OECD's \emph{AI Incidents Monitor} (\url{https://oecd.ai/en/incidents}) or the \emph{AI Incident Database} (\url{https://incidentdatabase.ai/}).} rely on news articles, data on non-public incidents may also be crucial for effective incident monitoring and risk-estimation \cite{obrien_coordinated_2024}. Cooperation on a common definition of an `AI incident' could lay the foundation for usefully interoperable domestic incident monitoring frameworks. Furthermore, any specification of protocols and best practices for incident reporting and monitoring would need to specify the information pertaining to an incident that should be shared with whom, including the type and magnitude of incident for which reporting should be required.

\textbf{Secure weight infrastructure standards.}
While the specific security measures individual AI development organisations use to protect AI model weights must remain confidential, there may be scope for international collaboration on developing standardised technical frameworks for securing weights without revealing the weights themselves or organisations' other confidential information. This could include joint research into cryptographic protocols specifically designed for protecting large-scale model weights, methods for third parties to verify appropriate security measures without gaining access to the weights themselves drawing on precedents such as nuclear safeguard inspections \cite{iaea_iaea_2014}, and interface specifications for hardware security modules designed for AI model protection \cite{kulp_hardware-enabled_2024, nevo_securing_2024}. Another precedent is the international collaboration on cryptographic standards through organisations like the ISO,\footnote{See, for example, the security requirements for cryptographic modules \cite{noauthor_isoiec_2012}.} that enabled countries to work together on security frameworks while maintaining their own secure implementations. However, such collaboration would need to be carefully structured to ensure that only high-level protocols and standards are shared and that the collaboration does not reveal vulnerabilities in existing security systems. For example, researchers have investigated cryptographic methods to secure AI model weights during inference and deployment using Zero-Knowledge Proofs \cite{kang_scaling_2022}, and hardware security modules (HSMs) have been proposed to protect AI models from unauthorised access and tampering \cite{pan_device-bind_2022}.

\textbf{Meta-analyses.}
May 2024 saw the publication of the interim International Scientific Report on the Safety of Advanced AI, an international effort led by the UK's Department for Science, Innovation and Technology that aimed to \textit{``drive a shared, science-based, up-to-date understanding of the safety of advanced AI.''} \cite{bengio_international_2024} The report represents a collaborative effort between 30 countries, including the US and China, as well as the EU and the UN. Similarly, the UN's High-Level Advisory Body on AI has recommended that the UN establish an \textit{``independent international scientific panel on AI''} which should issue \textit{``an annual report surveying AI-related capabilities, opportunities, risks, and uncertainties,''} among other actions \cite{un_hlab_govering_2024}.\footnote{A broader discussion of how these initiatives could inter-relate can be found in \cite{pouget_future_2024}.} These examples show that diverse states can, and furthermore do, coordinate on initiatives that aim to advance a shared technical and scientific understanding of AI, including the global risks that it may pose or exacerbate. Such `literature review'-style research efforts that establish a scientific basis for research funding and resource allocations, and more systematic meta analyses could aid in establishing greater alignment and enable better consensus building between rivals. However, it is worth noting that there may be a risk that such efforts become co-opted by political pressures to advance national interests.

\subsection{Infrastructure}
We use the term `AI safety infrastructure' to refer to systems and processes (be they hardware, software, organisational, or otherwise), external to an AI system, that facilitates research and development activities relating to AI safety. This could take the form of hardware infrastructure, such as computational resources, or software tooling such as code packages for conducting certain types of research \cite[see e.g.,][]{fiotto-kaufman_nnsight_2025}.
Cooperating on AI safety infrastructure could have large benefits for ensuring interoperability of ongoing research and development activities in different jurisdictions. This may allow for greater global distribution of the benefits of AI \cite{adan_voice_2024}, as well as facilitate further collaboration on empirical AI research.

\subsubsection{Risks of cooperation}

\textbf{Advances global frontier capabilities.}
Due to the broad, multi-purpose nature of many forms of infrastructure, there is a risk that the developments made through cooperation on infrastructure could be applied, perhaps with minor repurposing, to advance frontier AI capabilities. For example, general software packages aimed at facilitating AI safety research could potentially also facilitate more effective research and development that advances frontier AI capabilities. The marginal contribution of this infrastructure may however be minimal, due to the large and expanding amount of existing infrastructure for AI research and deployment.

\textbf{Differentially advances a rival's capabilities.}
Similarly to advancing global frontier capabilities, infrastructure could plausibly be used by rivals to facilitate their own strategic capabilities. However, since this infrastructure will be available to all cooperating parties and thus make it less costly for a given party to detect a rival's progress toward a decisive strategic capabilities on a shared system, it is less likely that this will confer a significant \emph{differential} advantage to a rival that goes undetected.

\textbf{Exposes other sensitive information.}
To the extent that developing some forms of shared infrastructure for AI safety builds upon existing national infrastructure, doing so may require providing sensitive details regarding existing infrastructure to adversaries. This could raise a broad range of national security-relevant risks. However, many forms of infrastructure, such as high-level evaluations frameworks such as the UK AI Safety Institute's \emph{Inspect}\footnote{\url{https://inspect.ai-safety-institute.org.uk/}} or Singapore's AI Verify Foundation's \emph{Project Moonshot},\footnote{\url{https://aiverifyfoundation.sg/project-moonshot/}} likely do not carry this risk.

\textbf{Provides opportunity for motivated actors to take harmful action.}
Infrastructure, being broadly multi-purpose and applicable to many potential downstream applications and uses, may be particularly susceptible to misuse by malicious actors. For example, compute resources, while valuable for running large-scale safety research experiments, can also be used to develop systems for intentionally harmful purposes if suitable safeguards are not put in place. Additionally, infrastructural developments may be vulnerable to tampering on the part of a collaborator, for example through the deliberate insertion of backdoors.

\subsubsection{Subareas}

\textbf{Infrastructure for evaluating systems.}
The development of infrastructure for conducting evaluations could aid in promoting an open ecosystem of research and evaluation into the capabilities and safety of AI systems.\footnote{See existing initiatives e.g. Inspect (\url{https://inspect.ai-safety-institute.org.uk/}) or Project Moonshot (\url{https://aiverifyfoundation.sg/project-moonshot/}).} In particular, standardising software infrastructure and methods for evaluating AI systems (perhaps facilitated by the open-sourcing of codebases) can aid in providing a shared understanding of the capabilities and risks of systems. It could also enhance interoperability, allowing countries to better interpret and build on the evaluations performed in other countries. While sharing future evaluation methods may help advance adversaries' capabilities by providing a clear metric for improvements or disclosing techniques to elicit upper-bound capabilities, cooperating on evaluation infrastructure would likely not incur these same risks.

\textbf{Building an international research cluster.}
Some areas of cutting-edge AI safety research demand a large amount of computational resources due to the immense size of the models involved \cite{besiroglu_compute_2024}. Some states are thus starting to invest in public compute infrastructure to be made available to academic researchers to catalyse domestic AI research, including on AI safety.\footnote{See e.g., \url{https://nairrpilot.org/}.} If a multilateral partnership between states is to collaborate on AI safety research, it may be necessary for the partnership to jointly develop and maintain shared computational resources. Additionally, a shared cluster could incorporate many of the verifiable computing suggestions covered in the previous subsection, as a proof-of-concept for their utility. This could bring the benefit of mutual trust in the computational infrastructure underlying shared research projects, and serve to pool resources to achieve efficiencies of scale, as in other large international research projects such as CERN or the International Space Station.

\textbf{Automatic control and supervision systems.}
Alongside the evaluation of models, it is also important to evaluate the systems and infrastructure in which these models are deployed and controlled. The development of robust monitoring and control systems for AI is becoming increasingly critical as alignment and robustness properties of models remain fragile. Despite significant advances in alignment techniques such as RLHF, models remain vulnerable to jailbreaking \cite{casper_open_2023}. While there is a growing industry focused on AI monitoring solutions,\footnote{See, for example, \emph{LLM Guard} (\url{https://github.com/protectai/llm-guard}) \emph{Azure AI Content Safety} (\url{https://azure.microsoft.com/en-us/products/ai-services/ai-content-safety}), and \emph{Lakera Guard} (\url{https://www.lakera.ai/lakera-guard}).} current AI safety approaches offer limited guarantees and may not generalise well to more agentic systems or novel failure modes. The \emph{`Benchmarks for the Evaluation of LLM Safeguards'} \cite{dorn_bells_2024} represents an initial effort to systematically evaluate monitoring systems. There remains a need for more comprehensive evaluation frameworks that can assess monitoring systems' effectiveness against a broader range of threats, including cyber-attacks, potential biohazards, and intentional subversion by models \cite{greenblatt_ai_2024, griffin_games_2024}. Cooperating on the evaluation of control and supervision systems does not seem to incur the same risks as cooperating on the direct evaluation of models, including potentially advancing a rival's strategic AI capabilities.

\subsection{Evaluation methodologies}
Methods and resources for reliably evaluating the capabilities and safety of AI systems, for example through benchmarking, red-teaming, human uplift studies, or agent evaluations, have become a centrepoint of AI (safety) research, particularly by governments \cite{feffer_red-teaming_2024, uk_ai_safety_institute_early_2024, mouton_operational_2023}. Cooperation on such methods could ensure interoperability, enabling jurisdictions to share and build upon each other's assessment results -- creating a more efficient global system for AI evaluation. Global cooperation in particular could be especially useful for evaluating language models in multilingual settings, due to the varied language and culture expertise necessary to robustly evaluate such systems \cite{romanou_include_2024}.

\subsubsection{Risks of cooperation}
\textbf{Advances global frontier capabilities.}
As evaluations are predominantly about assessing a system's capabilities or safety, rather than making improvements in these areas, cooperation on AI evaluations are unlikely to directly advance the global frontier. Indirect effects may still occur, such as benchmark-chasing, whereby the existence of a challenging evaluation becomes a target, spurring increased effort in order to improve a system's score on that evaluation. However, this could itself be harnessed for benefit in the case of predominantly safety-focused evaluations.

\textbf{Differentially advances a rival's capabilities.}
Some evaluation methodologies specify elicitation techniques aimed at extracting upper-bound performance from systems on dangerous or dual-use tasks. For example, the instruction to language models to `think step by step' generated significantly improved performance on many tasks \cite{wei_chain--thought_2023}, and is now incorporated directly into the system prompt of some reasoning models, as well as being a valuable technique for evaluations.
Sharing elicitation techniques could thus be particularly sensitive due to the potential for direct application for improving a system's capabilities. However, cooperation aimed at improving the efficacy or robustness of evaluation techniques for non-sensitive system capabilities may not pose this risk. A separate concern is that having access to quantitative evaluation methodologies for potentially dangerous system capabilities could provide 
a guiding measure for an actor trying to develop a system with those capabilities.

\textbf{Exposes other sensitive information.}
Depending on the focal area of the evaluation, cooperation may require the sharing of sensitive information, for example if assessing a model's cyber-offensive capabilities. However, there are also many domains where this would not be the case, for example assessing a model's propensity to generate false or biased content. By focusing cooperative efforts on these latter evaluation subjects, rivals can avoid having to reveal sensitive, domain-specific information. Early proofs-of-concept suggest that it is also possible to create secure evaluation environments that preserve the privacy of test data \cite{trask_secure_2024}.

\textbf{Provides opportunity for motivated actors to take harmful action.}
The degree to which cooperating on developing evaluations could allow motivated actors to take harmful actions depends to a large extent on the form that cooperation takes. For example, jointly developing benchmarks or other evaluation datasets seems to confer limited risk. However, more involved joint testing exercises may necessitate providing collaborators with privileged access to evaluation systems and infrastructure, as well as the models being tested, both of which could potentially be compromised.

\subsubsection{Subareas}

\textbf{Ensuring the reliability of evaluations.}
Recent research has highlighted challenges in ensuring the reliability of AI evaluation methodologies \cite{chang_survey_2024}. For example, concern has been raised regarding how overlap between a model's training and test datasets can artificially inflate scores on benchmarks and other evaluations \cite{zhang_language_2024, zhou_dont_2023}. Cooperation could focus on methods for improving the reliability of current approaches to AI evaluation.

\textbf{Advancing the science of evaluations.}
Finally, research cooperation could aim to advance the science of AI evaluation -- that is a more rigorous understanding of the theoretical basis of AI evaluation \cite{burden_evaluating_2024}. Since such a research agenda would be more foundational in nature, it is less likely that there will be a significant risk of leaking sensitive information regarding contemporary AI systems, or providing opportunities to cause harm.

\section{Conclusion}
Geopolitical rivals often have incentives to cooperate on strategic technologies, for example to address risks posed by that technology which span across national borders. However, such cooperation itself can pose risks which must be managed if the benefits of cooperation are fully realised.
In this paper we have provided an overview of current international cooperation on AI in an important case of geopolitical rivalry, and outlined four sources of risk pertinent to cooperation on technical AI safety that are under-addressed in current risk mitigation strategies. 
Based on this, we assess the extent to which these risks may be realised by four areas of technical work in AI safety which have been suggested as potential areas of international cooperation, finding that development of verification mechanisms and protocols may be well-suited for cooperation.
Future research could aim to extend the analysis of this paper by considering additional areas of technical AI safety research, propose more concrete policy recommendations concerning international cooperation on AI, or consider how states' domestic technology policy issues could relate to international cooperation.
We hope that this paper can serve as a foundation for further international cooperation efforts to address the risks associated with AI. 

\begin{acks}
The authors would like to thank Alan Chan, Lujain Ibrahim, Brodi Kotila, Benjamin Prud'Homme, and three anonymous reviewers for insightful comments and feedback.

AC acknowledges funding from a UKRI AI World Leading Researcher Fellowship (Grant EP/W002949/1) and from a JPMC Research Award.
\end{acks}

\newpage
\bibliographystyle{ACM-Reference-Format}
\bibliography{mybib}

\newpage
\appendix

\section{Summary table of candidate areas of cooperation}
Table \ref{tab:Candidate_areas} provides a summary of the candidate areas for cooperation, as described in section \ref{sec:candidate_areas}.

\begin{table}[h!]
    \small
    \centering
    \begin{tabular}{rp{3.2cm}p{3.2cm}p{3.2cm}p{3.2cm}}
         \toprule
         & \textbf{Advances global frontier capabilities} & \textbf{Differentially advances a rival's capabilities} & \textbf{Exposes other sensitive information} & \textbf{Provides opportunity for harmful action} \\ \midrule
         & \emph{Minimal} & \emph{Minimal} & \emph{Minimal/moderate} & \emph{Minimal/moderate} \\
         \rotatebox[origin=r]{90}{\textbf{Verification}}
         & Verification largely aims to attest to claims made about a given system/process, rather than improve it. While some areas of verification research could uncover new properties that advance system capabilities, these can be avoided when cooperating.
         & While verification methods may involve revealing a set of  model properties that need to be verified, the risk from this is typically low as the properties to be verified can be restricted to those known to all parties in advance.
         & The development of verification technologies, such as hardware-enabled mechanisms, could require disclosing sensitive information
         However, methods that are independent of the specification of the AI system and related infrastructure are also being developed.
         & When rivals jointly develop verification systems, they may insert hidden backdoors to fake compliance. While the difficulty of concealing such backdoors varies, this risk can be managed through open-source development and bug bounties to help detect subversion attempts. \\ \midrule
         & \emph{Minimal} & \emph{Minimal} & \emph{Minimal} & \emph{Minimal} \\
         \rotatebox[origin=r]{90}{\textbf{Protocols}}
         & Developing protocols is a process of codifying existing knowledge, rather than extending the knowledge frontier.
         & Cooperation on protocols development could focus on areas where all parties have shared knowledge. Thus no expertise would be shared that could allow a rival to advance their capabilities.
         & Developing protocols and standards usually only draws upon broadly-known knowledge. 
         Any areas requiring national security sensitive information could be avoided.
         & Protocol development does not involve any direct manipulation of AI systems, though standardisation has sometimes been used to advance unilateral interests
         \\ \midrule
         & \emph{Minimal/moderate} & \emph{Minimal/moderate} & \emph{Moderate} & \emph{Moderate} \\
         \rotatebox[origin=r]{90}{\textbf{Infrastructure}}
         & Infrastructure, being multi-purpose, could be useful, though not critical, in facilitating capabilities advances.
         & Infrastructure, being multi-purpose, could be instrumental, though not critical, in allowing a rival to advance their capabilities. Though the use of shared infrastructure could increase the likelihood that such applications are detected by cooperating parties.
         & To the extent that developing shared infrastructure builds upon existing infrastructure, doing so may require divulging sensitive information regarding this existing infrastructure.
         & Infrastructure's multi-purpose nature makes it vulnerable to misuse -- compute resources meant for beneficial AI research could be repurposed for harmful aims, and collaborative infrastructure projects risk tampering through deliberately inserted backdoors. \\ \midrule
         & \emph{Minimal/moderate} & \emph{Minimal/moderate} & \emph{Moderate} & \emph{Minimal/moderate} \\
         \rotatebox[origin=r]{90}{\textbf{Evaluations}}
         & Evaluations concern the assessment of a system’s capabilities rather than their improvement. While dangerous capability evaluations may specify capability elicitation techniques, the impact of this on the overall capabilities frontier is minimal compared to dedicated efforts to improve model capabilities.
         & Evaluation methods are closely tied to techniques for eliciting upper-bound capabilities from AI systems. Collaboration on evaluations may thus allow a rival to advance their capabilities through having a better awareness or understanding of such techniques.
         & Evaluations in sensitive areas (such as CBRN/Cyber) can involve a large amount of specialist domain knowledge. Collaborating on evaluations in these domains risks the transfer of this sensitive knowledge.
         & 
         Sharing benchmark methodologies is unlikely to provide significant opportunity for harm. However, joint testing wherein a rival receives access to a given model could provide greater opportunities for harm, although the chance that this happens completely undetected is still low. \\
         \bottomrule
    \end{tabular}
    \caption{Preliminary assessment of the risks of cooperating on the four technical AI safety areas discussed.}
    \label{tab:Candidate_areas}
\end{table}

\end{document}